\documentclass[12pt,preprint]{aastex}

\slugcomment{paper draft to ApJ (10/17/2005)}

\shorttitle{CO Mapping of EP Aqr}
\shortauthors{J. Nakashima}

\begin{document}


\title{BIMA CO Observation of EP Aqr the Semiregular Pulsating Star with a Double Component Line Profile}


\author{Jun-ichi Nakashima\altaffilmark{1}}
\affil{Department of Astronomy, University of Illinois at Urbana-Champaign, 1002 W. Green St., Urbana, IL 61801}
\email{junichi@asiaa.sinica.edu.tw}

\altaffiltext{1}{Current Address: Academia Sinica Institute of Astronomy and Astrophysics, P.O. Box 23-141, Taipei 106, Taiwan}


\begin{abstract}
This paper reports the results of a Berkeley-Illinois-Maryland array interferometric observation of EP Aqr, a semiregular pulsating star with a double component line profile in the CO $J=1$--0 line. The broad component shows a flat-top profile, and the narrow component shows a spiky strong peak. Though the previous single dish observations suggested that the CO $J=2$--1 line exhibits a Gaussian-like profile, the CO $J=1$--0 line does not. The spatial distributions of both the narrow and broad components appears to be roughly round with the same peak positions. No significant velocity gradient is seen. The spatial-kinetic properties of the molecular envelope of EP Aqr are reminiscent of a multiple shell structure model rather than a bipolar flow and disk model. A problem of this interpretation is that no evidence of interaction between the narrow and broad component regions is seen. A Gaussian-like feature seen in the CO $J=2$--1 line might play a key role to understand the spatio-kinetic properties of the molecular envelope of EP Aqr.
\end{abstract}




\section{INTRODUCTION}

Asymptotic giant branch (AGB) stars occasionally exhibit a molecular line profile with very small line widths less than 5 km s$^{-1}$, much smaller than a typical AGB outflow velocities. Such a narrow line profile is usually superimposed on a broad pedestal component (this profile is often called a ``double component profile'', or a ``composite profile''), but is sometimes independently found without the broad component \citep{kna98,ker99,win03}. A non-negligible number of AGB stars exhibit a narrow line; in fact, the narrow line has been detected from 5--10\% of a sample of AGB stars \citep{win03}. Narrow line profiles have been detected toward a wide variety of AGB stars. For instance, \citet{kna98} reported narrow lines detected toward both M- and C-type AGB stars; several semiregular pulsating stars also exhibit narrow lines \citep{ker99}. 

AGB stars exhibiting a narrow line profile are notable in some respects. First, chemically unusual AGB stars occasionally exhibit a narrow line profile. For instance, \citet{kah96} reported double component profiles in the CO radio lines of BM Gem and EU And; these stars are ``silicate carbon stars,'' which simultaneously have a carbon-rich (C-rich) central star and oxygen-rich (O-rich) circumstellar material. \citet{nak04b} have found a double component profile in radio molecular spectra of the SiO maser source with a rich set of molecular species, IRAS 19312+1950, and have suggested that this object is identified as an O-rich AGB (or post-AGB) star \citep{nak00,nak04a,nak04b,nak05b}. Second, according to recent observations, some of AGB stars exhibiting a narrow line profile might harbor a Keplerian rotation disk. \cite{ber00} have reported a tentative detection of such a disk in the O-rich AGB star with a double component profile, RV Boo. Similarly, the existence of a Keplerian rotation disk has been observationally suggested in X Her, which is also an O-rich AGB star with a double component profile \citep{nak05a}. Though a physical relationship between narrow lines, chemical peculiarity and a Keplerian rotation disk is still unclear, \citet{kah96} have suggested that narrow lines found in silicate carbon stars might be explained by gravitationally stable material in the form of a distorted or puffed-up slowly rotating disk, in which O-rich material is trapped. In contrast, to explain a double component profile, \citet{kna98} have advanced a hypothesis that takes multiple shell structure into account, in which each shell has different expanding velocities produced by episodic mass loss with highly varying gas expansion velocities. 

EP Aqr is a semi-regular variable (SRb) of period $\sim 55$ days (this period is uncertain) at a distance of 135 pc; its spectral type is M8III \citep{per97}. \citet{dum98} evaluated the stellar effective temperature at 3236 K. \citet{win03} and \citet{kna98} have produced high velocity resolution spectra of the CO $J=1$--0, $J=2$--1 and $J=3$--2 lines. All three lines show a double component profile with a narrow component ($V_{\rm exp} \sim 1.4$ km s$^{-1}$) superimposed on a broad pedestal component ($V_{\rm exp} \sim 11$ km s$^{-1}$), both centered at $V_{\rm lsr} \sim -34$ km s$^{-1}$. The mass loss rates derived from these CO observations are $2.3 \times 10^{-7}$ M$_{\odot}$ (broad component) and $1.7 \times 10^{-8}$ M$_{\odot}$ (narrow component). \citet{ker99} have reported the same value for the center velocity ($\sim -33.7$ km s$^{-1}$) of the two CO $J=1$--0 components. However, \citet{olo02} obtained slightly lower expansion velocities, 1.0 and 9.0 km s$^{-1}$, through a line-fit accounting for turbulence. EP Aqr has also been detected in the SiO ($v=0$, $J=2$--1) and ($v=0$, $J=3$--2) thermal emission by \citet{gon03}. Both lines are centered at $V_{\rm lsr} \sim -32$ km s$^{-1}$ and have a width corresponding to $V_{\rm exp} \sim 8 $km s$^{-1}$. Both lines have spectral profiles with only one component. 

This paper reports the results of a high-angular-resolution CO observation of EP Aqr with the Berkeley-Illinois-Maryland (BIMA) array. The main objectives of this research are (1) to investigate spatio-kinematic properties of the circumstellar molecular envelope of EP Aqr with a high-angular-resolution observation, and (2) to compare the spatio-kinetic properties with those of other AGB stars with a double component profile (i.e., X Her and RV Boo). As a result, the spatio-kinetic properties of EP Aqr are found to be very different from those of X Her and RV Boo.


\section{OBSERVATION AND RESULTS}
Interferometric observations of the CO $J=1$--0 line at 115.271202 GHz in EP Aqr were made with the BIMA array from 2004 January to March. The instrument was described in detail by \citet{wel96}. The array used 10 elements in 3 configurations (B, C and D). The observations were interleaved every 25 minutes with the nearby point sources, P2134+00 and 2148+069, to track the phase variations over time. The absolute flux calibration was determined from observations of Uranus and MWC349, and is accurate to within 20\%. The final map has an accumulated on-source observing time of about 12.3 hr. Typical single-sideband system temperatures range from 250 to 500 K. The velocity coverage is about 350 km s$^{-1}$, using three different correlator windows with a bandwidth of 50 MHz each. The velocity resolution is 1.03 km s$^{-1}$. The phase center of the map is R.A.$=21^h$46$^m$31.849$^s$ , decl.$=-2^{\circ}$12$'$45.923$''$ (J2000). Data reduction was performed with the MIRIAD software package \citep{sau95}. Standard data reduction, calibration, imaging, and deconvolution procedures were followed. Robust weighting of the visibility data gives a $5.15'' \times 3.68''$ CLEAN beam with a position angle of 10.4$^{\circ}$. The rms noise per 1.0 km s$^{-1}$ is 1.5$\times$10$^{-1}$ Jy beam$^{-1}$.

Figure 1 shows a spatially averaged spectrum of EP Aqr in the CO $J=1$--0 line. The averaged area is a circle with a diameter of 15$''$ centered at the phase center. A double component profile is clearly seen in the spectrum as reported in previous papers \citep{kna98,win03}. The broad pedestal component shows a flat-top profile, and the narrow component shows a strong spiky peak. \citet{win03} pointed out that the line profile of the narrow component in the $^{12}$CO $J=2$--1 line appears to be Gaussian; however, we do not have the spatial resolution to confirm this. The narrow component peaks at $V_{\rm lsr}=-34$ km s$^{-1}$, and its expansion velocity is calculated to be 2.1 km s$^{-1}$ by fitting a parabola. The expansion velocity of the broad component is 10 km s$^{-1}$; this is a half of a maximum width of the line at the zero intensity. The central velocity of the broad component is $-34$ km s$^{-1}$. These line parameters are consistent with previous values. In comparison with the single dish spectrum taken by the SEST 15m telescope \citep{ker99}, roughly 90\% of the single dish flux is recovered in the present interferometric observation. To check the continuum emission, frequency channels were integrated over 150 MHz using the opposite sideband to the line observation. The upper limit of the continuum flux at the 3 mm band (center frequency $\sim$112 GHz) is $1.9 \times 10^{-2}$ Jy.

The left panel of Figure 2 shows velocity integrated intensity maps in the CO $J=1$--0 line; for comparison, the right panel shows similar maps of X Her in the CO $J=1$--0 line, which also has a double component line profile \citep[the X Her maps are taken from][The recovered rate of the IRAM single dish flux has been confirmed to be more than  95\% --- this was incorrectly given on 5\% by Nakashima 2005]{nak05a}. The thick and thin contours represent the blue- and redshifted wings of the broad component, respectively; the lowest contours correspond to a 5$\sigma$ level, which was measured in an emission-free region in the images. The dashed gray contours represent the narrow component. Usage of contours is common in the right and left panel. At a glance, the overall structure of the molecular envelope of EP Aqr, which is clearly resolved by our synthesized beam, is significantly different from that of X Her. As reported by \citet{nak05a}, X Her exhibits a bipolar flow (corresponding to the thick and thin contours) and a possible disk (corresponding to the dashed contour). In contrast, all the thick, thin and dashed contours in the left panel appear to be roughly round, and the positions of the intensity peaks of the three components correspond within a beam size. Interestingly, the three components exhibit almost the same spatial sizes, though the narrow component is slightly elongated in the north-south direction. The diameter of the round structure seen in the left panel of Figure 2 is about 15$''$ corresponding to $3.0 \times 10^{16}$ cm at the distance of 135 pc. For better understanding of the spatio-kinetic properties, a position--velocity diagram is presented in Figure 3. Because the narrow component region is slightly extended in the north--south direction, one of the cuts used for Figure 3 (cut A) is taken along with the north--south elongation of the narrow component region, and another cut (cut B) is taken in the perpendicular direction. No significant velocity gradients are seen at these or other position angles. In Figure 3 the spatial size of the broad component region (represented in the vertical axes) is apparently smaller than that of the narrow component; however, this is due to a low signal-to-noise ratio in the corresponding velocity range. In fact, the size of the broad component region seen in the velocity integrated intensity map (see Figure 2) is almost the same with that of the narrow component region.


\section{DISCUSSION}
Until now, two AGB stars with a double component profile have been observed with radio interferometers in the CO $J=1$--0 line \citep[i.e., RV Boo and X Her,][]{ber00,nak05a}. These are all O-rich semiregular pulsating stars. However, the spatio-kinetic properties of EP Aqr are clearly different from those of X Her and RV Boo as seen in Section 2. If this is real, AGB envelopes with a double component line profile include, at least, two types of kinematics. As stated in Section 1, two hypotheses have been proposed to explain a double component profile: a slowly rotating puffed-up disk with a bipolar flow \citep{kah96} and double shell structure, in which each shell has different expanding velocities produced by episodic mass loss with highly varying gas expansion velocities \citep{kna98}. Previous observations suggested that the spatio-kinetic properties of X Her and RV Boo are possibly explained by the disk and bipolar flow model. In contrast, the properties of EP Aqr is rather reminiscent of the multiple shell model, because no significant velocity gradient is seen in position-velocity diagrams, and also because the overall structure of the CO emission region is roughly round rather than bipolar. One might consider a possibility that the properties of EP Aqr are explained by a disk and bipolar flow model if we assume a pole-on view; however, such a case is unlikely.

A problem in the interpretation as multiple (double) shell structure is that we cannot see any evidence of physical interaction between the narrow and broad component regions in the present observation. According to the spatial sizes and expanding velocities, the dynamical time scales of the narrow and broad component regions are roughly estimated to be 1270 yr and 160 yr, respectively; these time scales are similar to that of a typical AGB circumstellar envelope. The similar spatial scales but different time scales show that physical interaction must take place, and we see no observational evidence for this. For example, if material of the narrow component region is swept up by that of the broad component, the swept up material should be seen as a ring-like feature in the left panel of Figure 2. As pointed out in Section 2, the narrow component of the CO $J=2$--1 line detected in the previous single dish observation exhibits a Gaussian profile; in contrast, the CO $J=1$--0 line does not show such a Gaussian profile. This difference might potentially mean that an interacting region between the narrow and broad component regions cannot be easily detected in the CO $J=1$--0 line: if the interacting region has high temperature, it might not be easily seen in the CO $J=1$--0 line. Furthermore, if the interacting region has a high turbulence velocity caused by the interaction, the Gaussian profile seen in the CO $J=2$--1 line might be produced by the turbulence. (Of course, the absence of the Gaussian wing in the CO $J=1$--0 line might be just a signal-to-noise issue.) To check this conjecture, CO mapping observations in the CO $J=2$--1 line are required.


\section{SUMMARY}
This paper has reported the results of BIMA observations of a semiragular pulsating star with a double component line profile, EP Aqr, in the CO $J=1$--0 line. The main results are:
\begin{enumerate}
\item A double component profile is clearly seen in the spectrum as reported in the previous observations. The broad component shows a flat-top profile, and the narrow component is spiky strong peak. A Gaussian profile, which has been seen in the $J=2$--1 line, is not confirmed in the CO $J=1$--0 line.

\item Intensity maps of both the narrow and broad components are roughly round and centered at the same position. No significant velocity gradient is seen in the position-velocity diagrams.

\item The spatial-kinetic properties of EP Aqr is rather reminiscent of a multiple shell model suggested by \citet{kna98}. A problem with this interpretation is that no evidence of physical interaction between the narrow and broad component regions are seen in the present observation. A Gaussian-like wing feature seen only in the CO $J=2$--1 line might play a key role to understand this problem.
\end{enumerate}


\acknowledgments
The author thanks Shuji Deguchi for useful discussions. The author also thanks Jill Knapp, whose comments and suggestions as a referee was very helpful to improve the manuscript. This research has been supported by the Laboratory for Astronomical Imaging at the University of Illinois and by NSF grant AST 0228953, and has made use of the SIMBAD and ADS databases.

\begin{figure}
\epsscale{.60}
\plotone{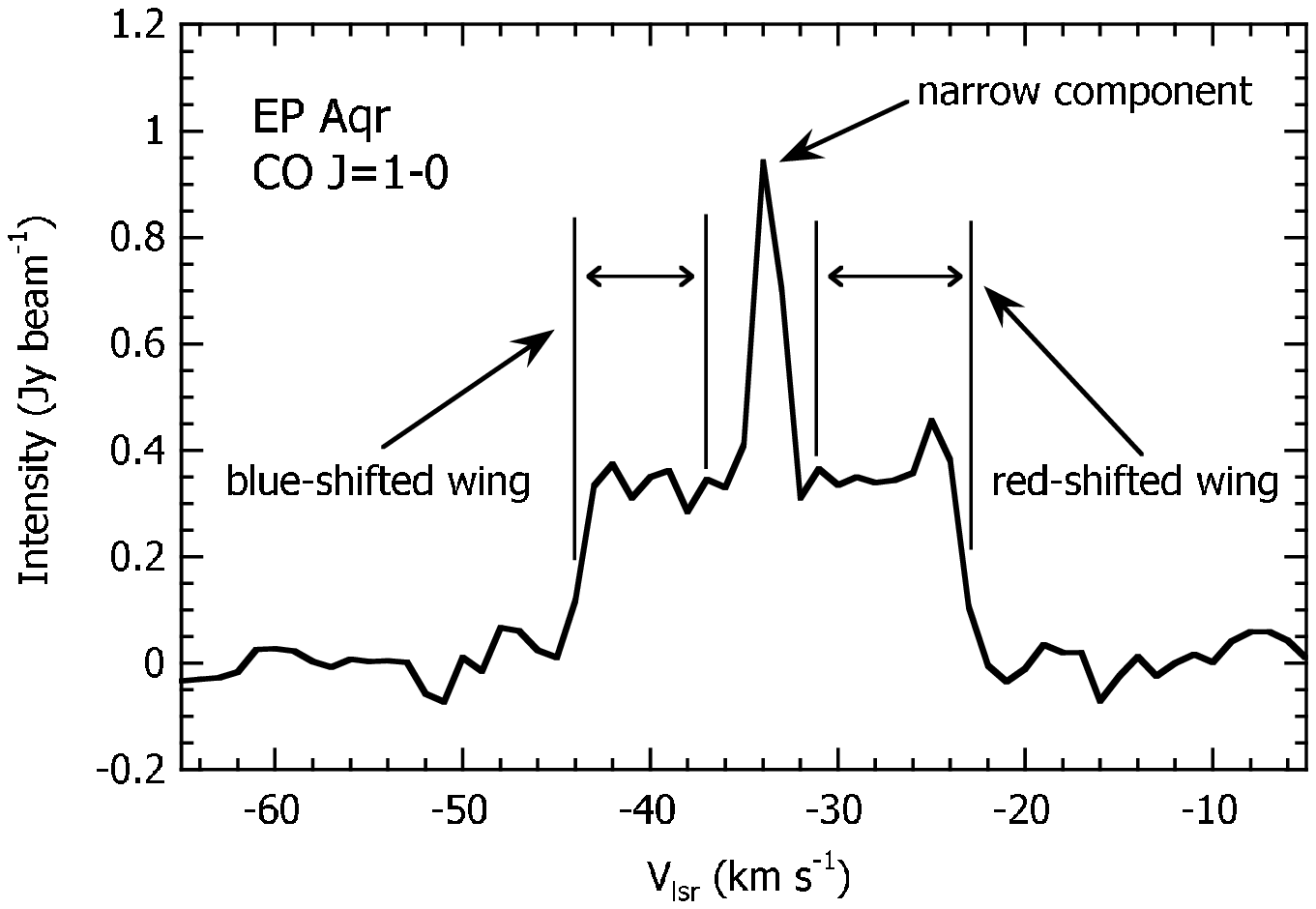}
\caption{Spatially averaged spectrum of EP Aqr in the CO $J=$1--0 line. The averaged area is a circle with a diameter of 15$''$; the averaging circle is centered on the mapping center. The vertical solid lines represent the velocity ranges selected for the blue- and red-shifted halves of the broad component. Each component is indicated. \label{fig1}}
\end{figure}

\clearpage

\begin{figure}
\epsscale{.80}
\plotone{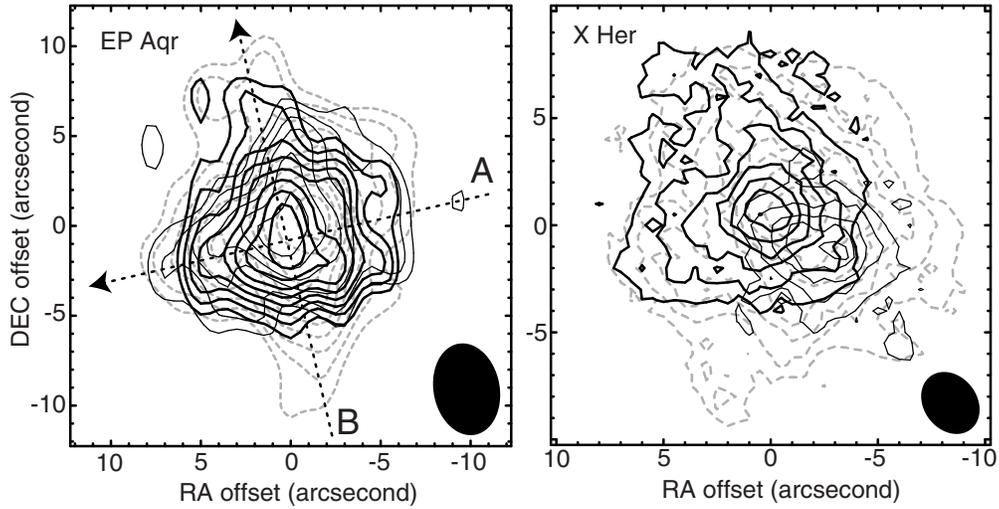}
\caption{{\it Left panel}: Velocity-integrated intensity maps of EP Aqr. The thick and thin contours map the blue- and redshifted wings of the broad component, respectively; the gray dashed contours map the narrow component. The synthesized beam size is indicated in the lower right corner. The contours start at the 5 $\sigma$ level, with increment every 2 $\sigma$. The 1 $\sigma$ levels for the thick, thin, and dashed contours are $5.1 \times 10^{-2}$, $5.1 \times 10^{-2}$, and $8.9 \times 10^{-2}$ Jy beam$^{-1}$. The velocity integration ranges for the blue- and redshifted wings are $-44$ to $-37$ and $-31$ to $-23$, respectively; the width of the integration range of the narrow component is 3 km s$^{-1}$ (the peak velocity is taken at the center of the range). The dotted arrows represent the cuts used for Fig. 3. {\it Right panel}: Similar map of X Her \citep[taken from][for comparison]{nak05a}. The contours are the same as in the left panel \citep[see][in details]{nak05a}. \label{fig2}}
\end{figure}

\clearpage

\begin{figure}
\epsscale{.60}
\plotone{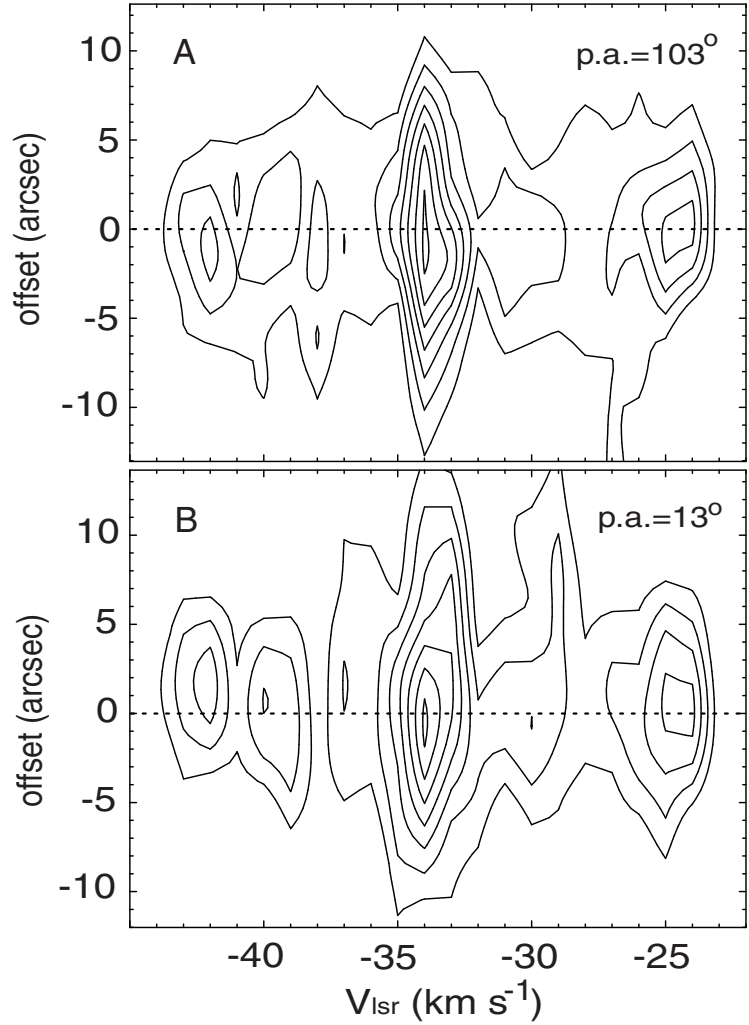}
\caption{Position-velocity diagrams for the cuts indicated in Fig. 2. The contours start at the 2 $\sigma$ level, with increment every 1 $\sigma$. The 1 $\sigma$ level is $1.5 \times 10^{-1}$ Jy beam$^{-1}$. The names of the cuts and the position angles indicated in the upper left and upper right corners of each panel, respectively. The dashed horizontal lines represent the origin of the offset axes. \label{fig3}}
\end{figure}

\clearpage

\end{document}